\documentclass[12pt]{article}
\usepackage{fleqn}

\usepackage{graphicx}
\usepackage[figuresright]{rotating}

\newcommand{\AmS}{{\protect\the\textfont2
  A\kern-.1667em\lower.5ex\hbox{M}\kern-.125emS}}

\title{New experimental data for the decays $\phi\to\mu^+\mu^-$ and 
$\phi\to\pi^+\pi^-$ from SND detector. }
\author{
M.N.~Achasov, V.M.~Aulchenko, K.I.~Beloborodov, A.V.~Berdyugin,\\
A.V.~Bozhenok, D.A.~Bukin, S.V.~Burdin, T.V.~Dimova,
S.I.~Dolinsky, \\ A.A.~Drozdetsky, V.P.~Druzhinin, M.S.~Dubrovin,
I.A.~Gaponenko, \\
V.B.~Golubev, V.N.~Ivanchenko, P.M.~Ivanov, A.A.~Korol, \\ S.V.~Koshuba, 
G.A.~Kukartsev, I.N.~Nesterenko, A.V.~Otboev, \\
E.V.~Pakhtusova, V.M.~Popov, A.A.~Salnikov,
S.I.~Serednyakov, \\ V.V.~Shary, Yu.M.~Shatunov, V.A.~Sidorov, 
Z.K.~Silagadze,\\ 
Yu.V.~Usov, A.V.~Vasiljev, A.S.~Zakharov\\
{\em Budker Institute of Nuclear Physics and}\\
{\em Novosibirsk State University, Novosibirsk 630090, Russia}\\
{ ~}\\
{talk given by S.~Burdin} \\ {at 8th International Conference on
   Hadron Spectroscopy (HADRON 99),}\\{ Beijing, China, 24-28 Aug 1999}
}
\date{}
\begin{document}
\maketitle
\begin{abstract}
   The processes $e^+e^- \to \mu^+\mu^-$ and $e^+e^- \to \pi^+\pi^-$
 have been studied with SND
detector at VEPP-2M $e^+e^-$ 
collider in the vicinity of $\phi(1020)$ resonance. 
The branching ratios 
$B(\phi\to\mu^+\mu^-)=(3.30\pm 0.45\pm 0.32)\times 10^{-4}$ and 
$B(\phi\to\pi^+\pi^-)=(0.71\pm 0.11\pm 0.09)\times 10^{-4}$
were obtained.
\end{abstract}

\section{Introduction}

  The both decays $\phi\to\mu^+\mu^-$ and $\phi\to\pi^+\pi^-$ give 
interference patterns in the energy dependences of the cross sections 
of the processes 
\begin{eqnarray}
\label{mumu}
& &  e^+e^-\to\mu^+\mu^-, \\
\label{pipi}
& & e^+e^-\to\pi^+\pi^-
\end{eqnarray}
in
the region of $\phi$ resonance. The interference amplitude is
determined by the branching ratio of the corresponding decay. The
table value of the branching ratio
$\mathrm{BR}(\phi\to\mu^+\mu^-)=(2.5\pm0.4)\times10^{-4}$ \cite{PDG}
is based on the experiments in photoproduction
\cite{Hayes}. In $e^+e^-$ collisions one can measure
the leptonic branching ratio of $\phi$ meson
$B_{e\mu}=\sqrt{B(\phi\to\mu^+\mu^-)\cdot B(\phi\to e^+e^-)}$.
Such measurements were performed in Orsay \cite{Orse} and Novosibirsk
\cite{Chil}, but their accuracy was not too high.

  The experimental result on the decay $\phi\to\pi^+\pi^-$ \cite{PDG}
  does not agree well with the theoretical predictions (see for
  example \cite{Ach}), but the improvement of the accuracy is needed.

  Data collected at VEPP-2M \cite{vepp2m} with SND detector
  \cite{SND} in the vicinity
  of $\phi$ resonance allow to improve the accuracy of the
  measurements of these decays.
  The result on the decay $\phi\to\mu^+\mu^-$ \cite{my}
 was obtained using 1996 data sample with
 the total integrated 
 luminosity $2.61$~pb$^{-1}$ and corresponding number of 
 $\phi$ mesons about $4.6\times10^{6}$. The 1998 data
  were used to measure the decay $\phi\to\pi^+\pi^-$. During 1998
 about $13.2\times10^{6}$ $\phi$ mesons were produced with the  
 integrated luminosity $8.6$~pb$^{-1}$. 

\section{Event selection}

\begin{figure}[htb]
\begin{minipage}[t]{75mm}
\includegraphics[width=70mm]{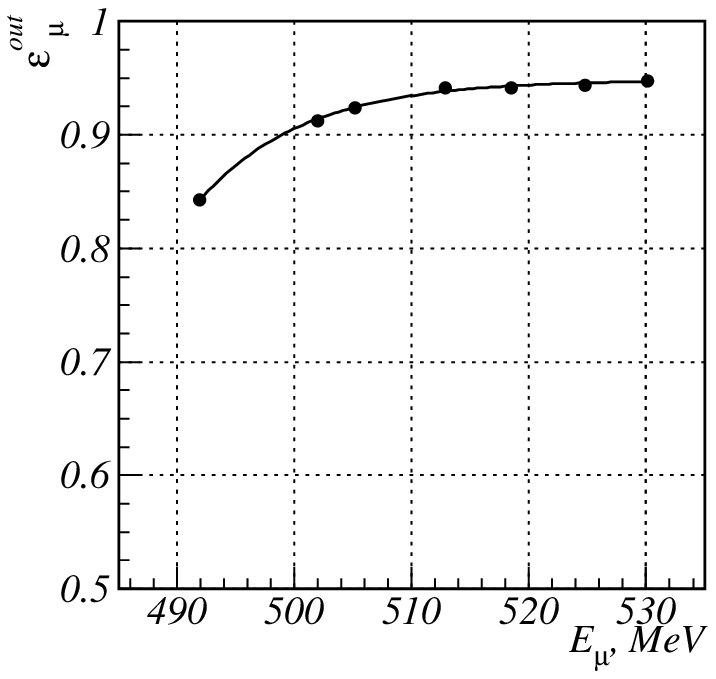}
\caption{ The energy dependence of the probability to  hit the outer system 
for muons.}
\label{fig:promu}
\end{minipage}
\hspace{\fill}
\begin{minipage}[t]{75mm}
\includegraphics[width=70mm]{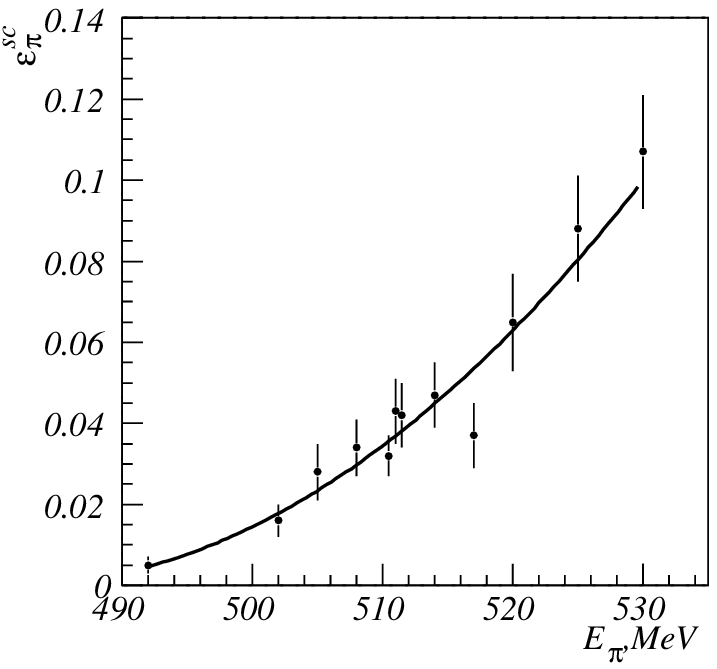}
\caption{The energy dependence of the probability to  hit the outer system 
for pions. }
\label{fig:propi}
\end{minipage}
\end{figure}

  The preliminary selections were
  the same for the both processes.
 The events with two collinear
  charged tracks were selected. The cuts on the angles of
  acollinearity in azimuth and polar directions were following: 
$\mid\Delta\varphi\mid<10^\circ$, $\mid\Delta\theta\mid<25^\circ$.
To suppress the beam and cosmic background the production point of charged
particles was required to be within $0.5$~cm from the
interaction point in the azimuth plane and $\pm 7.5$~cm along
the beam direction (the longitudinal size of the interaction region
 $\sigma_{z}$ is about $2$~cm).
The polar angles of the charged particles were limited in the range
$45^\circ<\theta<135^\circ$, corresponding to the acceptance angle of
the outer system \cite{SND}. The outer system allows to distinguish 
between the processes (\ref{mumu}) and (\ref{pipi}) 
due to the difference in the probability to hit the outer system
  for  muons and  pions. These
  probabilities are shown in Fig. \ref{fig:promu} and \ref{fig:propi}.
The data were divided for two samples (``muons'' and ``pions'')
 in dependence on the existence of the hit in the outer system.
The admixture of the events of the process (\ref{pipi}) in the sample
 ``muons'' was about 3\% at $\phi$-resonance energy. The process
 (\ref{mumu}) gave 15\% contribution to the sample ``pions'' 
at the same energy. 

 Other sources of background are the cosmic muons and the process
$ e^+e^-\to e^+e^-$. 
The cosmic background was significant 
 only for the process (\ref{mumu}) and was suppressed by the time of
flight system, which is a part of the outer system. 
The events of the process $ e^+e^-\to e^+e^-$ did not contribute
to the sample ``muons'' due to the suppression by the outer system. 
To reduce the background from these events for the process (\ref{pipi}) 
 the procedure of $e/\pi$ separation based 
on the energy depositions in the calorimeter layers was used. 
The events of the process $ e^+e^-\to e^+e^-$ were suppressed by factor
 $3.6\times10^{-4}$, while only 7\% of the events of the process under
 study were lost. 
The remaining background from the process $ e^+e^-\to e^+e^-$ was
about 1.5\% at $2E_b=1020$~MeV.

 The processes $e^+e^-\to\phi\to\pi^+\pi^-\pi^0$
 and $e^+e^-\to\phi\to K_SK_L$ gave resonance background to the
 process $e^+e^-\to\pi^+\pi^-$. 
The events of these decays were
 suppressed by restrictions on the energy depositions in the
 calorimeter layers. The contribution of the resonance 
background remained at the level of 0.5\% at $2E_b=1020$~MeV. 
 Because such background changes the visible interference
 pattern, 
 special efforts were made to subtract it. The selected events were
 divided in two parts by the cut on the parameter
 $\Delta\varphi$: $\mid\Delta\varphi\mid<5^\circ$ and
 $\mid\Delta\varphi\mid>5^\circ$. 
 To obtain the interference parameters
 the events from the first part were used.
 The resonance background was determined from the second part, where 
 its level is comparable with the visible cross section of the process 
(\ref{pipi}). The relationship between the quantities of the resonance
 background in two parts was calculated by Monte Carlo simulation
 \cite{unimod}.

 The detection efficiencies were obtained
 from the simulated data with using the experimental data for some
 corrections. The efficiencies for the processes (\ref{mumu}) and 
(\ref{pipi}) were 28\% and 12\% respectively 
  at the region of $\phi$ resonance.

\section{Data analysis}

  The energy dependence of the visible cross sections  
 of the processes under study 
 was fitted with the following formula:
$\sigma_{vis}(E)=\varepsilon(E)\sigma_{B}(E)\beta(E)+\sigma_{bg}(E)$,
where $\varepsilon$ -- the detection efficiency, $\sigma_{B}$ --
the Born cross section of the appropriate process, $\beta$ -- factor
taking into account the radiative corrections \cite{Kur1,Kur2}, 
$\sigma_{bg}$ is the cross section of the background. The Born cross
section was factorized into non-resonant and resonant parts: \\
$\sigma_{B}(E)=\sigma_{nr}(E)\mid Z\mid^2$,
$Z=1-Q\cdot e^{i\psi}\frac{m_\phi\Gamma_\phi}
{m^2_\phi-E^2-iE\Gamma(E)}$, 
where $Q,\psi$ -- modulus and phase of the interference amplitude,
$m_\phi,\Gamma_\phi$ -- mass and width of $\phi$ meson. Non-resonant
cross section was taken in the form
$\sigma_{\mu\mu}=83.50(\mathrm{nb})\frac{m^2_\phi}{E^2}$ for the
process (\ref{mumu}). A second order polynomial was used to describe the
cross section $\sigma_{nr}$ of the process (\ref{pipi}).  

The interference amplitude is related to the branching ratio of the
decay $\phi\to\mu^+\mu^-$ by the following expression: 
$Q_\mu=\frac{3\cdot\sqrt{B(\phi\to\mu^+\mu^-)\cdot B(\phi\to e^+e^-)}}{\alpha}$,
where $\alpha$ is the fine structure constant. For the decay
$\phi\to\pi^+\pi^-$:
$Q_\pi=\frac{6\cdot\sqrt{B(\phi\to\pi^+\pi^-)\cdot B(\phi\to e^+e^-)}}
{\alpha\beta_\pi^{3/2}(m_\phi)\mid F_\pi\mid}$,
where $F_\pi$ -- the pion form factor at the maximum of 
$\phi$ resonance, $\beta_\pi=(1-4\cdot
m^2_\pi/E^2)^{1/2}$. 

\begin{figure}[htb]
\begin{minipage}[t]{75mm}
\includegraphics[width=70mm]{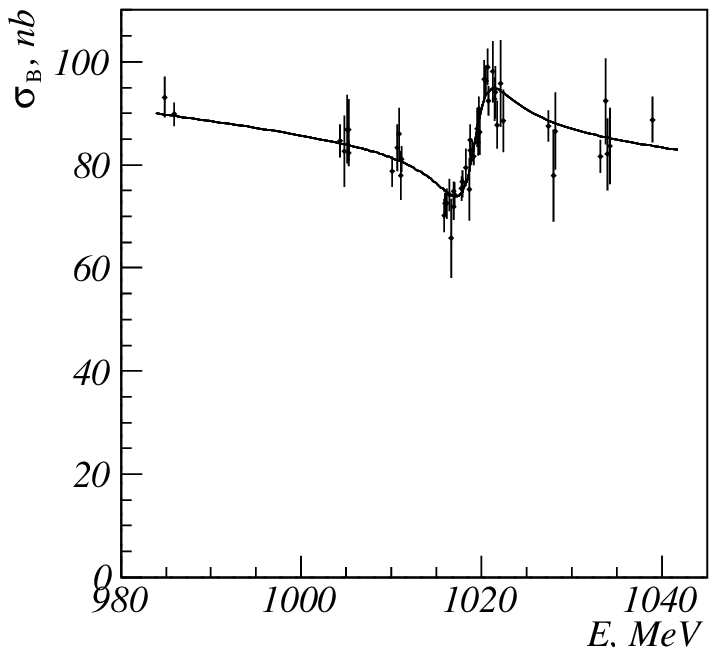}
\caption{ The Born cross section $\sigma_{B}$ of the process
$e^+e^-\to\mu^+\mu^-$.} 
\label{figmu}
\end{minipage}
\hspace{\fill}
\begin{minipage}[t]{75mm}
\includegraphics[width=70mm]{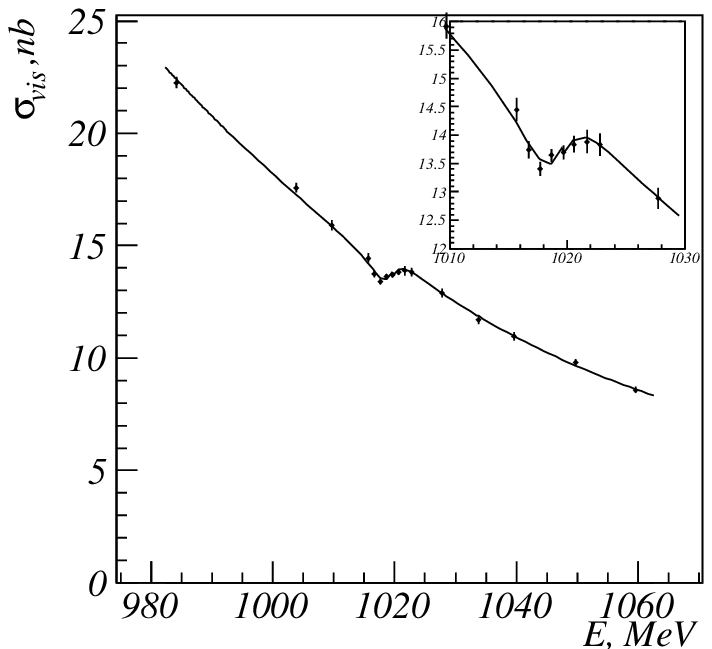}
\caption{The visible cross section $\sigma_{vis}$ for the process
$e^+e^-\to\pi^+\pi^-$. }
\label{figpi}
\end{minipage} 
\end{figure}

Fig. \ref{figmu} shows the cross section $\sigma_{B}$ of the process
(\ref{mumu}). The cross section $\sigma_{vis}$ for the process
(\ref{pipi}) is shown in Fig. \ref{figpi}.
The fit gave the following results: 
$Q_\mu=0.129\pm0.009$, $\psi_\mu=(4.5\pm3.4)^\circ$,
$\mid F_\pi\mid^2=2.98\pm0.02$, $Q_\pi=0.073\pm
0.005$, $\psi_\pi=-(34\pm4)^\circ$. The systematic errors are not
specified in these results.
From here the leptonic branching ratio of $\phi$ meson
$B_{e\mu}=(3.14\pm0.22\pm0.14)\times 10^{-4}$ and the branching
ratios  $B(\phi\to\mu^+\mu^-)=(3.30\pm 0.45\pm
0.32)\times 10^{-4}$,  $B(\phi\to\pi^+\pi^-)=(0.71\pm 0.11\pm
0.09)\times 10^{-4}$ 
were obtained. The real and imaginary parts of the interference
amplitude of the decay $\phi\to\pi^+\pi^-$ are
following: $\Re Z_\pi=(6.2\pm0.5\pm0.5)\times10^{-2}$,  
$\Im Z_\pi=-(4.2\pm0.6\pm0.4)\times10^{-2}$.

\section{Conclusion}
The measured value of $B_{e\mu}$ and the
branching ratio $B(\phi\to\mu^+\mu^-)$ are in good agreement with the
table branching ratio  $B(\phi\to e^+e^-)=(2.99\pm0.08)\cdot 10^{-4}$
\cite{PDG}. The accuracy of the result for the decay
$\phi\to\mu^+\mu^-$ is comparable with the accuracy of the table value 
$B(\phi\to\mu^+\mu^-)=(2.5\pm0.4)\cdot 10^{-4}$ \cite{PDG}.

The measured value of
$B(\phi\to\pi^+\pi^-)$ agrees
with the table value $B(\phi\to\pi^+\pi^-)=(0.8_{-0.4}^{+0.5})\times
10^{-4}$  and has much better accuracy. There is a discrepancy 
between our result and the preliminary result of CMD-2 \cite{CMD-2}.
The measured $\Re Z_\pi$ is much
lower than VDM prediction with standard $\rho-\omega-\phi$
mixing \cite{Ach}. Such low real part can be 
explained by the existence of direct decay of 
$\phi$ to $\pi^+\pi^-$ or 
non-standard $\rho-\omega-\phi$ mixing. 

\section{Acknowledgement}
 The work is partially supported by RFBR (Grants No 96-15-96327, 99-02-17155,
 99-02-16815, 99-02-16813) and STP ``Integration'' (Grant No 274).


\begin{thebibliography}{9}
\bibitem{PDG} Review of Particle Physics. The European Physical Journal
C 3 (1998).
\bibitem{Hayes}
 S.Hayes et al., Phys. Rev. D, V.4 (1971) 899. 
\bibitem{Orse}
 J.E.Augustin et al., Phys. Rev. Lett. 30 (1973) 462.
\bibitem{Chil}
 I.B.Vasserman et al., Phys. Lett. B 99 (1981) 62.
\bibitem{Ach}
 N.N.Achasov, A.A.Kozhevnikov, 
 Inter. Jour. Mod. Phys. A, V.7, No.20 (1992) 4825.
\bibitem{vepp2m} G.M.Tumaikin, Proc. of the 10-th Int. Conf. on
High Energy Particle Accelerators. Serpukhov, 1977, vol.1, p.443.
\bibitem{SND} M.N.Achasov et al., hep-ex/9909015, submitted to NIM, Section A.
\bibitem{my} 
M.N.Achasov et~al., Phys. Lett. B, 1999, vol. 456, p. 304.
\bibitem{unimod}
  A.D.Bukin et al., Preprint BINP 90-93 (1990). 
\bibitem{Kur1}
  A.B.Arbuzov et al., 
hep-ph/9702262, JHEP 9710 (1997) 001
\bibitem{Kur2}
  A.B.Arbuzov et al., 
hep-ph/9703456, JHEP 9710 (1997) 006.
\bibitem{CMD-2}
 R.R.Akhmetshin et al., Preprint Budker INP 99-11 (1999).
\end{thebibliography}
\end{document}